\documentclass[aps,twocolumn,nofootinbib,superscriptaddress,amsfonts,floatfix,preprintnumbers]{revtex4-2} 
\usepackage{graphicx,amsmath,amssymb,amstext}
\usepackage{amssymb,amsbsy,amsfonts,amsthm,color,bm}
\usepackage{mathtools}
\usepackage{epsfig}
\usepackage{graphicx}
\usepackage{subfigure}
\usepackage{hyperref}
\usepackage{cancel}
\usepackage{balance}
\usepackage{booktabs}
\usepackage{diagbox,multirow}
\graphicspath{{Figures/}}

\usepackage{color}
\usepackage[dvipsnames]{xcolor}
\usepackage{soul}

\renewcommand{\eqref}[1]{Eq.~(\ref{#1})}

\newcommand{\Beq}{\begin{eqnarray}}
\newcommand{\Eeq}{\end{eqnarray}}

\def\lsim{\mathrel {\vcenter {\baselineskip 0pt \kern 0pt \hbox{$<$} \kern 0pt \hbox{$\sim$} }}}

\newcommand{\ldbbar}{\lambdabar_{\mathrm{dB}}}

\newcommand{\dd}{\mathrm{d}}

\newcommand{\pd}{\partial}
\begin{document}

\title{Relativistic dynamical friction in Kerr-Schild coordinates}

\author{Zipeng Wang}
\email{zipeng.wang@ub.edu}
\affiliation{Institut de Ciències del Cosmos (ICC),
Universitat de Barcelona, Martí i Franquès 1,
ES-08028, Barcelona, Spain}

\author{Francesca Chadha-Day}
\email{francesca.chadha-day@durham.ac.uk}
\affiliation{Institute for Particle Physics Phenomenology Department of Physics, Durham University Durham DH1 3LE, U.K.}

\author{Katy Clough}
\email{k.clough@qmul.ac.uk}
\affiliation{School of Mathematical Sciences, Queen Mary University of London Mile End Road, London, E1 4NS, UK}

\author{James Maxwell}
\email{james.l.maxwell@durham.ac.uk}
\affiliation{Institute for Particle Physics Phenomenology Department of Physics, Durham University Durham DH1 3LE, U.K.}

\preprint{IPPP/26/75}


\begin{abstract}
Simulations of the evolution of matter on static metric backgrounds allow us to estimate backreaction effects like dynamical friction at first order in the matter density. Previous simulations of dynamical friction have used boosted isotropic and quasi-isotropic Kerr coordinates, which simplifies the extraction of the relevant diagnostics but results in a more complicated implementation of the evolution. In particular, the lack of horizon penetrating coordinates means that high resolutions are needed to ensure convergence, and the boosted, spinning black hole metric cannot easily be written in a closed form, making it more costly to evaluate.
In this paper we describe how to reformulate the calculation of the diagnostics in the boosted Kerr-Schild frame for the non-spinning case. We verify that our implementation reproduces previous results quantifying dynamical friction in wave-like dark matter in the appropriate regimes. We discuss the subtleties of transforming diagnostic observables between frames in a 3+1 evolution and develop a general framework for this process.
\end{abstract}


\maketitle


\section{Introduction} \label{sec:intro}

Dynamical friction is a purely gravitational drag force resulting from the gravitational attraction between a perturber and overdensities that form in the medium around it. First described by Chandrasekhar~\cite{Chandrasekhar:1943ys} for a nonrelativistic Newtonian perturber moving through a cloud of noninteracting particles, dynamical friction has been extended to  fluids~\cite{Ruderman:1971,Rephaeli:1980,Ostriker:1998fa}), different geometries~\cite{Tremaine:1984,Muto:2011qv,Vicente:2019ilr}), and to include relativistic corrections~\cite{Syer:1994vr,Petrich:1989,Barausse:2007ph,Correia:2022gcs}. In the context of wave-like scalar matter (a candidate for dark matter, for a review see \cite{Hui:2021tkt}), dynamical friction was first computed for a nonrelativistic Newtonian perturber~\cite{Hui:2016ltb}, and then extended to include velocity dispersion~\cite{Lancaster:2019mde,BarOr:2019}, self-gravity~\cite{Annulli:2020ilw,Annulli:2020lyc} (see also~\cite{Wang:2021udl,Chowdhury:2021}) or self-interactions~\cite{Hartman:2020fbg,Boudon:2022dxi}, binary systems~\cite{Buehler:2022tmr, Tomaselli:2023ysb}, and relativistic cases~\cite{Traykova:2023qyv,Traykova:2021dua,Vicente:2022ivh}. The ability to model dynamical friction and verify analytic approximations is important for work to incorporate environmental effects into waveform models, which can then be used for data analysis of gravitational wave catalogies, as was done in e.g. \cite{Roy:2025qaa} for the case of scalar dark matter studied here.

The gauge freedom of general relativity allows the same physical black hole spacetime to be cast in different forms, some of which are more amenable to numerical or analytic analyses. In the case of numerical evolution of matter fields on a fixed black hole background, Kerr-Schild coordinates are usually preferred, as they are horizon penetrating and therefore fields evolve smoothly across the horizon. Isotropic Schwarzschild coordinates suffer from coordinate singularities at the horizon, such that field oscillations will tend to ``bunch up'' there. Simulations in such coordinates still work, but unfeasibly high resolutions can be required, depending on how close the physics of interest is to the horizon. Analytic results, on the other hand, tend to be quoted with reference to asymptotically Schwarzschild-like observers. Naturally both systems describe the same physics, and so it is in principle straightforward to convert between the two frames to calculate observable quantities. In practice, however, there is quite a lot of scope for error in identifying the different systems to be used for the simulation volumes, charges and fluxes, and for understanding which quantities are inherently frame-dependent (even if the result, once the frame is chosen, is not itself coordinate dependent). 

In this work we discuss how to reformulate the previous numerical work of \cite{Traykova:2023qyv}, which was performed in (a comoving, boosted form of) isotropic coordinates, into (a comoving, boosted form of) Kerr-Schild coordinates. This work sought to calculate the dynamical friction in wave-like dark matter scenarios, and use the simulations to validate analytic expressions, and was extended to Magnus-like effects in spinning black holes in \cite{Wang:2024cej}. As well as being horizon penetrating, the Kerr-Schild metric also offers the practical advantage that it can more easily be written in a closed form in the boosted and spinning case, making it less costly to evaluate at each timestep. We summarise the general coordinate transformation framework required and apply it to the specific case of a non spinning, boosted black hole as an example. To verify our simulations, we compare our results to the analytic expressions of \cite{Traykova:2023qyv}, which were validated in isotropic Schwarzschild coordinates, and explain the agreement and disagreement in different regimes. We highlight the effect of frame choices where relevant and clarifying some distinctions that were glossed over in previous work. We close with a suggestion for extensions and applications of the work. Throughout the paper we use geometrised units in which $G=c=1$ and a $(-,+,+,+)$ metric signature.


\begin{figure}[t]
\centering
\includegraphics[width=0.48\textwidth]{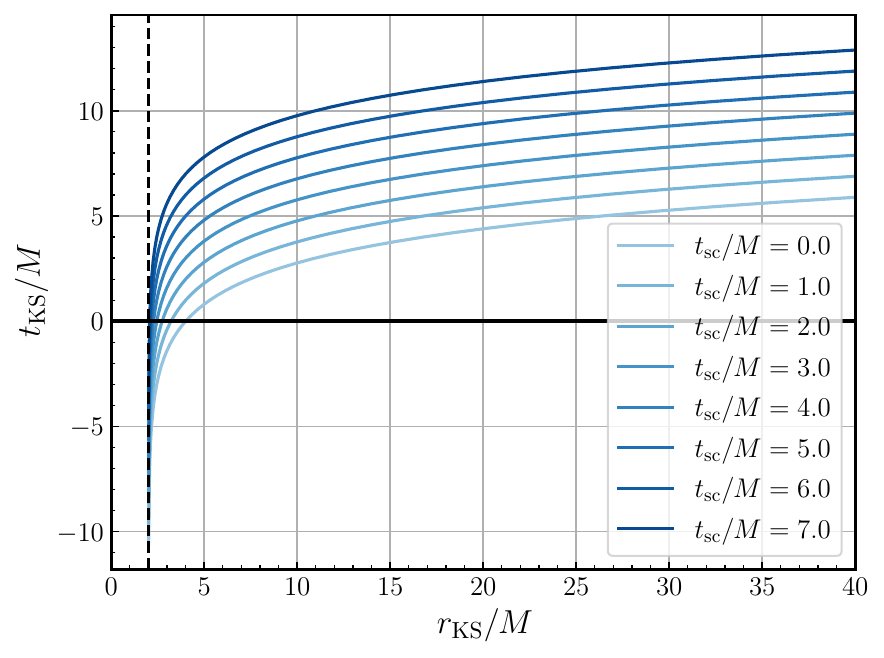}
\caption{Constant Schwarzschild time spatial slices in Kerr-Schild coordinates: Whilst the lines of constant $r$ are the same in both systems, the surfaces of constant $t$ are not aligned. The event horizon ($r=2M$) can be reached in a finite time in Kerr-Schild coordinates, but an infinite amount of time needs to pass in Schwarzschild coordinates to reach this point. We can also see that the asymptotic normal observers in Kerr-Schild only coincide with Schwarzschild ones in the limit of large $r$, and that the fall off in this discrepancy is slow. }\label{fig:ConstTimes}
\end{figure}

\section{Diagnostic formulation}
\label{sec:setup}

In this section we revise the general framework for measuring dynamical friction in numerical simulations on fixed metric backgrounds and discuss how to convert the diagnostics used in \cite{Traykova:2023qyv, Traykova:2021dua} into a different frame. In particular, we wish to transform between the diagnostic frame (DF), in which the measured charge (momentum) is defined, and the simulation frame (SF) that we use in our numerical simulation.

One potential issue is that the spatial hypersurfaces $\Sigma_R$ (assumed to be constant time spatial volumes within a fixed coordinate radius $r=R$) over which we measure the volume and surface integrals of the fluxes and charges will usually need to be defined in the SF coordinates, since doing otherwise would require us to reconstruct the full four dimensional spacetime from our 3+1 split. This is possible in theory but practically difficult and we would prefer to avoid it. The measurements in the DF and SF may therefore differ simply because the spatial hypersurface $\Sigma_R$ is a surface of constant SF time, which will not in general be a surface of constant DF time (see e.g. the KS and Schwarzschild frames illustrated in Fig. \ref{fig:ConstTimes}). That is, we are taking integrals of a dynamically evolving quantity over different 3D spatial volumes in a 4D spacetime, and so we have no reason to expect them to be the same. 

However, with the correctly defined charge, the dynamical friction measure should converge to the same solution in both slicings if the matter reaches a steady state within the volume, and if the DF and SF coordinate systems coincide outside the outer boundary of the spatial volume $r=R$ (this is true even if they do not coincide within it)\footnote{This follows from the observation in Landau and Lifshitz \cite{Landau:1975pou} that the conserved total charge (curvature plus matter), must be the same for any interior coordinate system, assuming the coordinate systems coincide outside of the volume.}. This will be approximately true for our dynamical friction simulations for Kerr-Schild as SF and Isotropic Schwarzschild as DF, where the coordinate systems approach the same slicing at large cloud radius $R$. At smaller $R$ we should expect some difference to arise because the frames no longer coincide at the boundary. 

The fluxes and charges themselves should be defined as those related to the translation generators of the DF coordinates that we want to use (then evaluated for the SF volume $\Sigma_R$). Specifically for momentum, we consider the charges related to infinitesimal translations of the DF coordinates $\xi^\mu=(\partial_{\bar{x}})^\mu$, where the components of the vectors field are $\xi^\mu=(0,1,0,0)$ when expressed in that frame. In general, these are not the same physical directions as translations of the SF coordinates $(\partial_{x})^\mu$, and so the components need to be transformed appropriately between frames. If we simply use the SF generator $(\partial_{x})^\mu$, then we measure a different physical charge.

We discuss these points and define more precisely the quantities to measure in the following sections.

\subsection{Relativistic DF force}

We are trying to measure the force on the BH due to the presence of an overdensity in the matter around it. In the Newtonian limit this would be characterised by the quantity
\begin{equation}
F_{Newton} = \frac{dP_{BH}^i}{dt} = \int_{\Sigma} \rho s^i dV ~,
\end{equation}
where $s^i$ is the vector separation between the BH and the volume element, $\rho$ is the local density, and $\Sigma$ is the 3d spatial volume of coordinate radius $R$ that contains the overdensity. 

There is no gravitational ``force'' in the relativistic picture, but we can think about the transfer of momentum from the matter to the curvature of the BH in the context of an asymptotically measured 4-momentum charge $\Theta_\mu{}^0$
\begin{equation}
    P_\mu =\int_{\Sigma_R} \Theta_\mu{}^0 ~ dV  ~.
\end{equation}
One can define such a charge for any volume assuming there is a conservation law of the form
\begin{equation}
    \partial_\nu \Theta_\mu{}^\nu = 0~.
\end{equation}
(We will see later that $\Theta_\mu{}^\nu$ is the Einstein complex defined in Sec.~\ref{sec:etensor}.)
The total $P_\mu$ will be made up of both matter and curvature contributions (as, for example, the ADM 4-momentum is).
Then for $\tau$ the proper time of the asymptotic observer
\footnote{This assumes that at the surface $\partial\Sigma_R$ a proper time can be meaningfully defined - at large $R$ in asymptotically flat space all observers on the surface will experience the same proper time $\Delta \tau$ in a coordinate interval $\Delta t$, but at a smaller fixed coordinate radius this may not be true. We will only consider the former case.}
the 4-force is
\begin{equation}
F^\mu_{Rel, curvature} = \frac{dP^{curvature}_\mu} {d\tau} ~.
\end{equation}
This assumes that we can in some way split $P_\mu$ into the curvature contribution and the matter contribution. Any such split will necessarily be coordinate dependent, since the stress-energy associated with curvature cannot be defined in a covariant way. Here we will use a prescription based on the Einstein pseudotensor, but this is not unique (one could also use the Landau-Lifshitz pseudotensor, for example).

\subsection{The Einstein pseudotensor}\label{sec:etensor}
The Einstein complex $\Theta_\mu{}^\nu$ is defined as 
\begin{equation}
\Theta_\mu{}^\nu
=
\frac{1}{16\pi}
\frac{\partial H_\mu{}^{\nu\sigma}}{\partial x^\sigma},
\end{equation}
where $H_\mu{}^{\nu \sigma}$ is the Freud superpotential
\begin{equation}
H_\mu{}^{\nu \sigma}
=
\frac{g_{\mu \rho}}{\sqrt{-g}}
\frac{\partial}{\partial x^\lambda}
\left[
-g\left(
g^{\nu \rho}g^{\sigma \lambda}
-
g^{\sigma \rho}g^{\nu \lambda}
\right)
\right],
\end{equation}
\begin{equation}
H_\mu{}^{\nu\sigma}=-H_\mu{}^{\sigma \nu}.
\end{equation}
The Einstein Complex can be shown to be made up of a matter and non-matter/curvature part,
\begin{equation}
\Theta_\mu{}^\nu
=
\underbrace{\sqrt{-g}\,T_\mu{}^\nu}_{\text{matter part}}
+
\underbrace{\sqrt{-g}\,t_\mu{}^\nu}_{\text{curvature part}} ,
\end{equation}
with the Einstein pseudotensor defined by combinations of first order derivatives of the metric
\begin{multline}
\sqrt{-g}\,t_\mu{}^\nu
=
\frac{1}{16\pi}
\left[
\delta_\mu{}^\nu
g^{\alpha\beta} 
\left(
\Gamma^\lambda{}_{\alpha\sigma}
\Gamma^\sigma{}_{\beta\lambda}
-
\Gamma^\lambda{}_{\alpha\beta}
\Gamma^\sigma{}_{\lambda\sigma}
\right) \right.
+ \\ \left.
g^{\alpha\beta}{}_{,\mu}
\left(
\Gamma^\nu{}_{\alpha\beta}
-
\delta_\alpha{}^\nu
\Gamma^\lambda{}_{\beta\lambda}
\right)
\right].
\end{multline}
The (coordinate) divergence of the Einstein complex is zero, that is $\partial_\nu \Theta_\mu{}^\nu = 0$. Therefore we obtain a conservation law
\begin{equation}
\underbrace{\partial_\nu (\sqrt{-g}\, T_\mu{}^\nu)}_{\text{matter part}}
+
\underbrace{\partial_\nu (\sqrt{-g}\, t_\mu{}^\nu)}_{\text{curvature part}} = 0.
\end{equation}
Note that this is not a tensorial equation and the split of what we call matter and curvature will very much depend on the chosen coordinate basis.
However, it is reassuring to note that the total charge so defined recovers the ADM momentum at large $R$ in an asymptotically flat, Cartesian basis, as
\begin{equation}
P_\mu^{\rm ADM}
=
\lim_{R\to\infty}
\int_{\Sigma_R}
\Theta_\mu{}^0 ~ dV
=
\lim_{R\to\infty}
\frac{1}{16\pi}
\int_{\partial\Sigma_R}
H_\mu{}^{0i}\,dS_i ~, \label{eq:P_ADM}
\end{equation}
making it a reasonable candidate for an equivalent finite radius quantity.

From this we define the 4-momentum charge related to the curvature as
\begin{equation}
P_\mu^{Rel, curvature} = \int_{\Sigma_R} \sqrt{-g} ~ t^0_\mu ~dV
\end{equation}
where the dV is the coordinate volume. 
The force is then
\begin{multline}
F_\mu = \left(\frac{dt}{d\tau} \right)\partial_0 \int \sqrt{-g} ~ t^0_\mu ~dV \\
= -\left(\frac{dt}{d\tau} \right) \left( \int \partial_j (\sqrt{-g} ~ t^j_\mu) ~ dV + \int \partial_\nu (\sqrt{-g} ~ T^\nu_\mu) ~ dV  \right) ~.
\end{multline}
Note that here we split time and space according to the coordinate directions, and the $\left(\frac{dt}{d\tau} \right)$ term again relates to the boundary observer's proper time. We can convert the first term on the RHS to a surface integral
\begin{equation}
\int \partial_j (\sqrt{-g} ~ t^j_i) ~ dV = \int \sqrt{-g} ~ t^j_i ~ dS_j
\end{equation}
where $dS_j$ is the coordinate surface element. Normally this term is zero (it is for the Kerr Schild and Schwarzschild metrics we consider), and is associated with the flux of GWs. This can usually be neglected relative to the leading order matter contributions (counting orders of the matter density $\rho M^2$ where $M$ is the black hole mass).  

\begin{figure*}[t]
\centering
\includegraphics[width=0.96\textwidth]{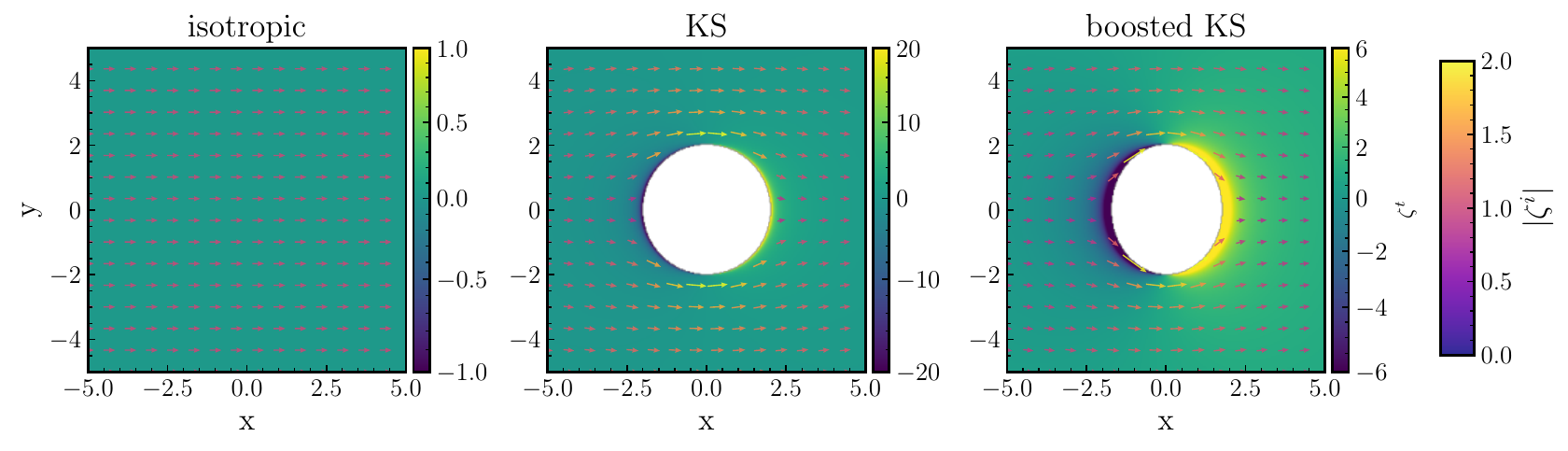}
\caption{The translation generator $\zeta^\mu$ in different systems of coordinates. The arrows show the direction and the magnitude for the spatial part of the vector, $\zeta^i$, and the colormap shows its time component, $\zeta^t$. In the isotropic coordinates (DF), $\zeta^\mu = (0,1,0,0)$, whereas it takes a more complicated form in the Kerr-Schild frame and the boosted Kerr-Schild (SF) frame. The central white region seen on the right two panels are not covered by the isotropic coordinates and hence has no $\zeta^\mu$ defined.  }\label{fig:vector_field}
\end{figure*}

It is shown in \cite{Clough:2021qlv} that what is called the source term $\mathcal{S}(\zeta)$ is related to the flux between the matter and curvature sectors
\begin{multline}
\int \sqrt{\gamma} ~ \mathcal{S}(\zeta) ~ dV = \int \sqrt{\gamma} \alpha ~T^\mu_\nu ~\nabla_\mu \zeta^\nu ~ dV 
\\
= \int \partial_\nu (\sqrt{-g} ~ T_\mu{}^\nu \zeta^\mu) ~ dV ~,
\end{multline}
where $\alpha$ and $\gamma$ are the lapse and shift of the ADM decomposition in the SF. The choice of vector field $\zeta^\mu$ defines the infinitesimal spacetime transformation associated with the charge of interest (e.g. an asymptotic spatial translation for linear momentum). In the coordinate system in which the charge is defined, the generator associated with translations along $x^a$ is
\begin{equation}
    \zeta_{(a)}^\mu
    =
    \left(\frac{\partial}{\partial x^a}\right)^\mu
    =
    \delta^\mu{}_a ~.
\end{equation}
In another coordinate system, the components of the same vector field must be transformed accordingly and will not, in general, remain equal to $\delta^\mu{}_a$.
In the coordinate system of the charge, 
\begin{equation}
S {}_{(a)} = - \int \partial_\nu (\sqrt{-g} ~ T^\nu{}_a) ~ dV ~.
\end{equation}
Therefore this source represents the flow of $a$-momentum from matter to curvature (or vice versa). The source quantity related to a particular transformation can still be evaluated in another frame, provided that $\zeta^\mu$ is transformed appropriately, and that the resulting terms involving $\partial_\mu \zeta^\nu$ are included when evaluating $\nabla_\mu \zeta^\nu$ (see Appendix \ref{sec:appendix} for the full expressions). 

In summary, in order to evaluate the flow of momentum from matter to curvature, we first identify the DF that we want to use to evaluate the charge (e.g. isotropic Schwarzschild), and specify the components of the $i$-momentum generator in that frame to be $\zeta^\mu=\delta^\mu_i$. For illustration, Fig.~\ref{fig:vector_field} shows $\zeta^\mu$ in the Schwarzschild isotropic coordinates (DF), the Kerr-Schild coordinates without the boost, and the boosted Kerr-Schild coordinates (SF). We then evaluate in any SF coordinate system (e.g. boosted Kerr-Schild)
\begin{equation}
\label{eq:f-extraction}
F_i = \left(\frac{dt}{d\tau} \right) \int \sqrt{\gamma} \alpha ~T^\mu_\nu ~\nabla_\mu \zeta^\nu dV
\end{equation}
where the prefactor $(dt/d\tau)$ is evaluated for the asymptotic observer in the DF.
One can show that in the non-relativistic limit, this quantity is precisely the analogue of the Newtonian integral.

\subsection{Excising the singularity}

As noted in \eqref{eq:P_ADM}, at large $R$ the finite volume expression approaches the ADM energy-momentum tensor, for which the momentum components are
\begin{equation}
P_{ADM}^i \equiv \frac{1}{8\pi} \int_{\partial\Sigma_\infty} (K^{ij} - K \gamma^{ij}) dS_j ~,
\end{equation}
where $K^{ij}$ is the extrinsic curvature of the spatial hypersurfaces. The quantity $P_{ADM}^\mu$ transforms in the usual special relativistic way with boosts etc when including the ADM energy as the time component of the 4-vector. In principle, we could use Gauss' law to convert this into a volume integral
\begin{equation}
P_{ADM}^i = \frac{1}{8\pi} \int_{\Sigma_\infty} (D_j K^{ij} - \gamma^{ij} D_j K) ~ dV ~,
\end{equation}
and then connect this (using the momentum constraint) to the momentum density of the spacetime in the volume
\begin{equation}
P_{ADM}^i = \int_{\Sigma_\infty} S^i~ dV ~,
\end{equation}
but we should be aware that in writing this we have implicitly assumed that the spacetime is regular within the volume, which it will not be if there is a BH. We may therefore need to add in the BH momentum by excising some small region around the BH and replacing it with a surface integral related to the curvature, as discussed in \cite{Clough:2021qlv}, as
\begin{equation}
\int_{\Sigma_{R_{inner}}} \sqrt{\gamma} ~ \mathcal{S}_i ~ dV = \int_{\partial \Sigma_{R_{inner}}} \sqrt{g} ~ T^j_i  dS_j ~.
\end{equation}
As shown in Fig 5 of \cite{Traykova:2021dua}, the total of the exterior volume and this surface integral is independent of the radius at which the excision is done.
\footnote{There is a lengthy discussion in Appendix A of \cite{Traykova:2023qyv} about whether one also needs to track the energy flux across this boundary, but it suffices to say that you do not, and if the above procedure is used, in which $\zeta^\mu$ is defined directly in the DF coordinates and not the SF ones, such corrections will be automatically accounted for in calculating the momentum fluxes.}

\subsection{Connection to flux at a finite radius}

In the case where the boundary of the volume is taken out to spatial infinity, a non-zero momentum curvature source must result in a change in the fluid momentum $S_i$, since the surface flux is equal to zero. 

At a finite radius boundary, we observe instead that at some point the fluid within the volume will reach a roughly steady state (i.e., the time derivative of $S_i$ is small), at which point the loss in momentum from dynamical friction is now balanced by a change in the net flux $J^i$ into the volume. So one will often see the statement that
\begin{equation}
F_i = \int_{\partial \Sigma_R} \sqrt{g} ~ J^j  dS_j ~.
\label{eq:AnalyticDF}
\end{equation}
This is taken to be the definition of dynamical friction in most analytic work, but it is really only true in the steady state; as we defined it, the exchange of momentum between curvature and matter is always the last ``source'' term, and in a general dynamical case it will be balanced both by an increase/decrease in the fluid momentum within the volume, and a flux to/from infinity. We could use \eqref{eq:AnalyticDF} in our simulations, but in general we find that it will take longer to settle to a stationary value than directly using the source term, especially for large volumes.

A final potential complication to note is that if the coordinate volume used for the evaluation of these quantities is moving, there is an advection term that must be added, as detailed in the appendix of \cite{Aurrekoetxea:2023jwk}.

\subsection{Choice of the diagnostic frame and $\zeta^\mu$}

In principle we can use translations of any frame to define its own momentum charge, and then measure a force on the black hole as the change in the curvature part of that charge. For example, we could also evaluate momentum-like quantities by setting the components of $\zeta^\mu = \delta^\mu_i$ in the boosted Kerr-Schild coordinates instead.

This has a number of disadvantages. The main one is that most analytic studies quote the force measured by the observer in the BH rest frame, for asymptotic observers for whom the time and normal directions of the hypersurface are coincident (i.e., in coordinates that do not contain a shift term). 
Therefore we should measure in an unboosted BH frame to agree to these results.
Another slightly more subtle point is that even in unboosted Kerr-Schild coordinates, the spatial hypersurfaces of constant $t$ are angled relative to the Isotropic Schwarzschild ones due to the radial shift term. This shift decays to zero at spatial infinity, so one might think that asymptotically one will recover the same result for both coordinate systems, and on a pointwise basis this is true. However, the decay is a relatively slow one, see Fig. \ref{fig:ConstTimes}, and therefore when integrated over a volume or surface, it still results in a spurious contribution in cases where the matter does not decay fast enough. In particular, in our dynamical friction case we start with a spatially constant $\rho$. Then since the difference in the translation generators $\delta\zeta^\mu\sim O(M/r)$ between the Kerr-Schild and Isotroptic asymptotic observers, our source term differs as
\begin{equation}
    \int T^\mu{}_\nu\nabla_\mu(\delta\zeta^\nu)\,dV
\sim\int^R \rho\,\frac{M}{r^2}r^2dr
\sim\rho MR ~.
\end{equation}
That is, it gives a contribution to the force that grows with $R$, even where initially density is constant and no wake has formed\footnote{In fact, it not only gives a contribution before any wake forms, it also gives a contribution with the wrong sign, as if there was a force that accelerated the BH. So a naive use of Kerr-Schild coordinates to define the translation gives an obviously unphysical result, if interpreted as the isotropic-frame drag force.}.
The two choices of $\zeta^\mu$ define different momentum charges. As a result, the force obtained using the untransformed Kerr–Schild coordinate generator should not be expected to equal the isotropic-frame dynamical-friction force.

In other cases, provided the asymptotic DF and SF generators approach each other fast enough compared to the fall-off of the matter (e.g. $\delta\zeta^\mu$ faster than $1/r^2$ for constant density), then using different coordinate systems to define the charge should recover the same result once a steady state is reached within the volume of interest.

We briefly note that in previous work, the diagnostic force was sometimes loosely referred to as the one ``measured by the coordinate observers''. This is not accurate, since the local contributions that are summed in the source volume integral are not those physically measured by the coordinate observers (even where SF and DF coincide).
A physical observer has a unit timelike four-velocity $u^\mu u_\mu=-1$. At each event, the observer’s instantaneous rest space consists of the vectors $s^\mu$ satisfying $u_\mu s^\mu=0$.
To measure momentum in some direction $i$, the observer chooses three unit vectors $e^\mu_{(i)}$ inside that rest space, satisfying 
\begin{equation}
 u_\mu e^\mu_{(i)}=0,
\qquad
g_{\mu\nu}e^\mu_{(i)}e^\nu_{(j)}=\delta_{ij}~,   
\end{equation}
which form the observer’s orthonormal tetrad. In a general curved spacetime, $e^\mu_{(i)}\neq \delta^\mu_i$. In particular, where the coordinate system includes a shift, a fixed-coordinate observer's ``space'' vector will generally include a coordinate time component. Even in the absence of a shift, the normalisation will result in a different locally measured charge to the one defined by an infinitesimal coordinate translation (the latter of which is used to define the global ADM charges, which is what we are concerned with here).

\subsection{Coordinate transformation framework}

In this subsection, we summarise how to calculate a particular force defined in a DF when simulated in a different SF. The force relates to a rate of change of a momentum charge as defined by some translation generator $\zeta^{\mu}$, specified in DF coordinates as $\zeta^{\mu}_{~(i)} = \delta^\mu_i$. We then define the current $J^{\mu}$ as $J^{\mu} = \zeta^{\nu} T^{\mu}_{\nu}$ \cite{Clough:2021qlv}, and this current can be calculated in any frame by transforming $\zeta^{\mu}$ to the SF coordinates appropriately. 

The quantity to be evaluated in the SF is then 
\begin{equation}\label{eq:summary}
F_i = \left(\frac{dt}{d\tau} \right) \int \sqrt{\gamma} \alpha ~T^\mu_\nu ~\nabla_\mu \zeta^\nu dV,
\end{equation}
which, in the steady state, equals the flux of the current $J^{\mu}$ corresponding to a DF spatial translation through the boundary surface of the simulation volume $V$. Note that as well as defining $\zeta^{\mu}$ in the DF, we must also include the proper time $\tau$ of observers who are stationary on the boundary of the volume as defined in the DF.

The integral for a 3+1D split defining $\alpha$ and $\gamma$ may then be performed in the SF. If the DF and SF coordinates coincide outside the outer boundary of the integration volume, $F^i$ as defined above will be identical in either coordinate system once the steady state is reached. Otherwise, there may be a difference corresponding to the fact that the spatial hypersurface of the SF is not a spatial hypersurface of the DF.

\section{Coordinate systems} 
\label{sec:coordinates}

The~$3+1$ ADM decomposition of the metric is given by
\begin{equation}
\dd s^2=-\bar\alpha^2\dd t^2+\bar\gamma_{ij}(\dd x^i + \bar\beta^i\dd t)(\dd x^j + \bar\beta^j\dd t)\,.
\end{equation}
In the DF, the lapse, shift, and nonzero components of the spatial metric are, respectively, 
\begin{equation}
\begin{gathered}
\bar \alpha = \frac{1-M/2\bar r}{1+M/2\bar r}\,,\qquad \bar\beta^i = 0\,, \\ 
\bar \gamma_{xx} = \bar\gamma_{yy} = \bar\gamma_{zz} = \left(1+ \frac{M}{2 \bar r}\right)^4  \,,
\end{gathered}
\end{equation}
where $\bar r^2  =  \bar x^2 + \bar y^2 + \bar z^2$ is the square of the radial coordinate in DF.

The Kerr-Schild metric is used as the SF, for which
\begin{equation}
\dd s^2
= \left(\eta_{\mu\nu} + 2 H\, l_\mu l_\nu\right)\dd x^\mu \dd x^\nu\,,
\end{equation}
where $\eta_{\mu\nu} = \text{Diag}\left( -1,1,1,1 \right)$, $H = M/r$, and $l_\mu = \left( 1, x/r, y/r, z/r\right)$. The SF coordinates are obtained by subjecting the above metric to a ``boost" in the ~$\pd/\pd\bar{x}$ direction, given by
\begin{equation}\label{eqn:boost_trans}
    t \rightarrow \gamma(t + v x) \quad
x \rightarrow x/\gamma \quad
y \rightarrow y \quad
z \rightarrow z.
\end{equation}
The result is a time-invariant metric in which the BH remains at the fixed coordinate position.
The ADM variables in the SF, $\alpha$, $\beta_i$, and $\gamma_{ij}$, can then be read off from the SF metric.

The final ingredient needed for our force extraction scheme is the coordinate transformation between the DF coordinates (DC) and the SF coordinates (SC). First, the coordinate transformation from the isotropic coordinates to the Kerr-Schild coordinates reads
\begin{align}\label{eqn:ISOtoKS}
    t_{\text{KS}} &= t_{\text{ISO}} + 2M\log\left| \frac{r_{\text{ISO}}}{2M}\left( 1 + \frac{M}{2 r_{\text{ISO}}} \right)^2 -1\right|\,,\nonumber\\
    r_{\text{KS}} &=  r_{\text{ISO}}\left( 1 + \frac{M}{2 r_{\text{ISO}}} \right)^2\,.
\end{align}

The Jacobian between DC and SC is obtained as
\begin{equation}
    J_{\text{DC}\rightarrow \text{SC}} = J_{\text{ISO}\rightarrow \text{KS}} \cdot J_{\text{boost}}^{-1}\,,
\end{equation}
where $J_{\text{boost}}$ is the Jacobian associated with the transformation \eqref{eqn:boost_trans}, $J_{\text{ISO}\rightarrow \text{KS}}$ is the Jacobian associated with \eqref{eqn:ISOtoKS} (in practice we translate these into relationships between Cartesian coordinates), and the dot stands for matrix multiplication.
The DC quantities, $\bar \alpha, \bar \beta_\mu$, and $\bar N_\mu$, are transformed into SC with $J_{\text{DC}\rightarrow \text{SC}}$, and the force extraction formula \eqref{eq:summary} is evaluated in SC.

As a final note, the velocity $v$ used in the metric and Jacobian calculations is not treated as a vector, but a mere parameter present in the coordinate transformation. This ensures the one-to-one correspondence between the DF and SF. This is possible because the value of $v$ for which the BH remains at fixed coordinates is the same in each frame i.e. $\frac{\partial t_{KS}}{\partial t_{ISO}} = 1$.

\begin{figure*}[t]
\centering
\includegraphics[width=0.96\textwidth]{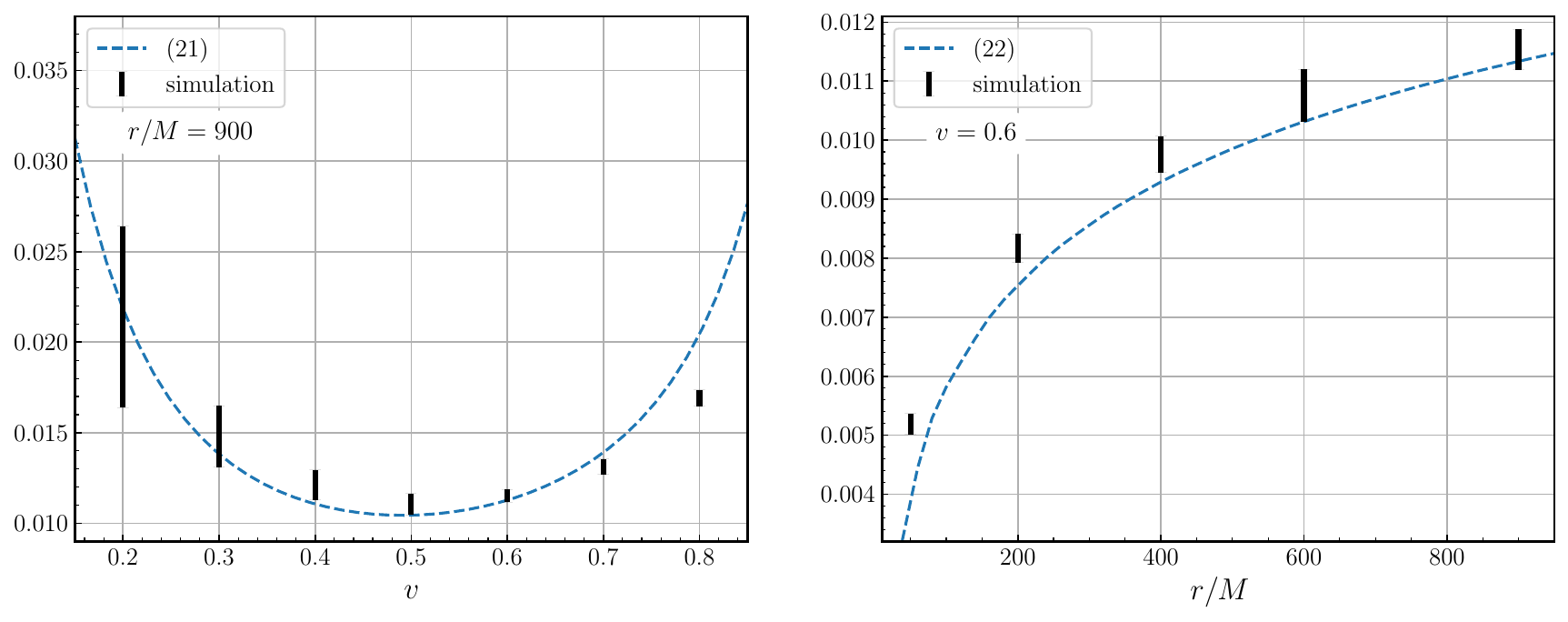}
\caption{Results of the dynamical friction force extracted with the updated extraction scheme compared with analytical formulae. The black bars show the ranges of the dynamical friction force extracted with \eqref{eq:f-extraction}. The left panel shows that, for a large cloud size of $r/M = 900$, we can recover the expected analytic relations in the relativistic regimes we study up to $v\sim0.7$. Above this the results diverge slightly but should agree for still larger $R$. The right panel shows that the simulation results deviate from the analytical result for small cloud sizes, due to differences in the slices of constant time used to evaluate the force. In all simulations displayed here, $\alpha/M = 0.05$.}\label{fig:Results}
\end{figure*}

\section{Analytic results for verification}
\label{sec:analytic}

We explore two scenarios from the wave dark matter dynamical friction paper of \cite{Traykova:2023qyv}. Both quantities are as measured by observers whose proper time is the coordinate time of the (unboosted) isotropic coordinates.

The first one applies for large cloud size $r$ and small scalar mass parameter $\alpha_s = \mu M$, for which
\begin{multline}
    F_{\bar{x}}\approx -\frac{4 \pi \rho M^2}{v^2} \gamma^2 (1+v^2)^2 \big\{ \overbrace{\ln \Lambda-1-{\rm Re} \Psi(1+i \beta)}^{\mathcal{D}(\bar{\alpha}_{\rm s}\ll 1, \Lambda\gg 1)} \\
    +\underbrace{\tfrac{4 v^3}{(1+v^2)^2}\tfrac{e^{\pi \beta}\pi \beta}{\sinh(\pi \beta)}}_{\mathcal{A}(\bar{\alpha}_{\rm s}\ll 1)\equiv \mathcal{A}_{\rm Unruh}}\big\}\,,\label{eqn:Fd_large-kr}
\end{multline}
where $\bar{\alpha}_{\rm s} = \gamma \alpha_{\rm s}$, $\rho$ is the asymptotic rest mass density of the medium, $\Psi$ is the digamma function,
\begin{equation}
    \beta \coloneqq \frac{(\frac{1+v^2}{v^2})M}{\ldbbar}= \bar{\alpha}_{\rm s} \left(\frac{1+v^2}{v}\right)\,,
\end{equation}
and
\begin{equation}
    \Lambda \coloneqq \frac{2 r}{\ldbbar}=2\gamma \mu v r\,.
\end{equation}
We expect to be able to recover this well from our SF evolution, for the reasons discussed above.

The second applies to smaller scalar field clouds (but still much larger than the BH size~($r/M\gg1$)) in the wave limit~$\beta \ll 1$, where we have:
\begin{multline}
    F_{\bar{x}}\approx -\frac{4 \pi \rho M^2}{v^2} \gamma^2 (1+v^2)^2 \big\{ \overbrace{{\rm Cin}( \Lambda)+ \tfrac{\sin \Lambda}{\Lambda} - 1}^{\mathcal{D}(\bar{\alpha}_{\rm s}\ll 1, \beta \ll 1)} \\
    +\underbrace{\tfrac{4 v^3}{(1+v^2)^2}\tfrac{e^{\pi \beta}\pi \beta}{\sinh(\pi \beta)}}_{\mathcal{A}_{\rm Unruh}}\big\}\,,\label{eqn:Fd_all-kr}
\end{multline}
where~${\rm Cin}(z) \coloneqq \int_0^z (1 - \cos t) \dd t /t$ is the cosine integral. 
We expect these values to disagree somewhat with our results, due to the fact that the slicings are physically different and do not coincide at the boundary for smaller $R$.

\section{Results}
\label{sec:results}

We extract the dynamical force in both the large-cloud and the small-cloud scenarios.
We stay in the wave regime $\bar \alpha_s \ll 1$ by choosing $\alpha_s M = 0.05$ in all our simulations. We also put the inner extraction boundary at $r=3M$. This choice is arbitrary and for small enough values the result does not depend on it \cite{Traykova:2021dua}.

In the first scenario, we set the outer extraction boundary at $R=900M$. This means $\Lambda \gg 1 $ and \eqref{eqn:Fd_large-kr} applies. In these simulations, the total size of the box is $4096M$, and a total of $9~(2:1)$ refinement levels are used. On the coarsest level the number of points is $N=128$, so the finest level used close to the BH has a resolution of $\Delta x = 0.0625M$. We also ensure at least 32 time steps per period of scalar oscillation throughout the domain. These conditions are the same as those used in the previous study \cite{Traykova:2023qyv}. 

The extracted magnitudes of the dynamical force are compared against the analytical expression, \eqref{eqn:Fd_large-kr}, in Fig.~\ref{fig:Results}. In the left panel we fix the size of the cloud to be $r = 900M$, and vary the speed of the black hole between $v=0.2-0.8$. In the right panel we fix the velocity to be $v=0.6$ and vary the radius at which the force is evaluated $R$ from $50M$ to $900M$.

In the case of large cloud sizes, the extracted force according to \eqref{eq:summary} agrees well with the analytical result \eqref{eqn:Fd_all-kr} up to $v=0.7$. At lower velocities, our results have larger uncertainties, as in the previous study (see the top panel of Fig.~1 in \cite{Traykova:2021dua}). However, we find that at $v=0.7$ and $v=0.8$, the simulation results deviate from \eqref{eqn:Fd_all-kr}. This is because the DF and SF integration surfaces enclose differently shaped regions due to the boost (that is, constant $r=R$ surfaces are not equivalent) and the difference reduces at larger $R$. Additionally, the boost compresses the length scale in the $x$ direction across the full grid, which means that a higher-resolution grid may be required in the most extreme cases. The challenge of resolving high boosts was encountered in previous studies \cite{Traykova:2021dua, Wang:2024cej}.

For small cloud sizes, $\Lambda \gtrsim 1$ the assumption of \eqref{eqn:Fd_large-kr} breaks down, but \eqref{eqn:Fd_all-kr} still applies. For $R=50M$, we set the total size of the box here at $512M$, and use a total of of $6~(2:1)$ refinement levels. On the coarsest level we still have $N=128$, so that on the finest level the resolution is $\Delta x = 0.0625M$ again. For the rest of the cloud sizes we keep the simulation setup as $L=4096M$ and $9$ refinement levels.

The dynamical friction force extracted from these simulations is shown in the right panel of Fig.~\ref{fig:Results}. We see the simulation results deviate from the analytical result more at smaller cloud sizes. As explained in Sec.~\ref{sec:setup}, the same dynamical friction force can only be recovered in the steady state by asymptotic observers who coincide outside the cloud volume $R$, so this is expected. We see that the difference decreases with the increase in the cloud size and the two results broadly agree at around $R\sim 600M$.

The agreement between the analytical expressions and the simulations presented here in the appropriate regimes confirm the validity of our force extraction scheme summarized by \eqref{eq:summary}. This enables us to extract the DF charge in another set of coordinates more suitable for numerical simulations. 

\section{Discussion}
\label{sec:summary}

In this work we set out a general framework for how to transform diagnostics defined in one fixed background frame to another in 3+1 numerical evolutions, and applied it to the case of a transformation between isotropic Schwarzschild and boosted Kerr-Schild coordinates for a non-spinning black hole, validating it against past results and analytic approximations for the appropriate regimes.

The computational cost of these simulations is similar to those in \cite{Traykova:2023qyv} for the non spinning case, since we did not investigate how much we could reduce the resolution and still recover the same results. In dynamical friction, most of the structure that contributes to the force is far from the horizon. This is not the case in other problems like the Magnus effect and superradiance, where structure forms closer to the horizon. 

In the future, the most useful application of this approach is to the case of spinning Kerr black holes in Kerr-Schild coordinates, in particular for calculations of the Magnus effect from dark matter around black holes, as was done in \cite{Wang:2024cej}. In this case the spinning Kerr-Schild metric is analytic, whereas the boosted quasi-isotropic version used in \cite{Wang:2024cej} did not have a closed form and some derivatives needed to be evaluated numerically from a metric stored on the grid. The use of Kerr-Schild coordinates should significantly reduce the cost and accuracy for a given resolution, which would allow us to push into more challenging regimes of high spins and high boosts, where the sign of the Magnus effect was found to reverse in some cases. We leave the study of such cases to future work.


\section*{Acknowledgements}

We would like to thank Josu Aurrekoetxea, Emanuele Berti, Pedro Ferreira, Thomas Helfer, Lam Hui, Dina Traykova and Rodrigo Vicente for helpful discussions, and the GRTL Collaboration (\url{www.grtlcollaboration.org}) for their development work on the GRDzhadzha code.

KC is supported by an STFC Ernest Rutherford fellowship, project reference ST/V003240/1, an STFC Research Grant ST/X000931/1 (Astronomy at Queen Mary 2023-2026), and the Simons Foundation International and the Simons Foundation through Simons Foundation grant SFI-MPS-BH-00012593-13. FCD and JM are supported by STFC under ST/X000745/1

This work used the DiRAC Memory Intensive service Cosma8 at Durham University, managed by the Institute for Computational Cosmology on behalf of the STFC DiRAC HPC Facility (\url{www.dirac.ac.uk}). The DiRAC service at Durham was funded by BEIS, UKRI and STFC capital funding, Durham University and STFC operations grants. DiRAC is part of the UKRI Digital Research Infrastructure. 
We also acknowledge the EuroHPC Joint Undertaking for awarding this project access to the EuroHPC supercomputer LUMI-C, hosted by CSC (Finland) and the LUMI consortium through a EuroHPC
Large Scale Access call.

For the purpose of Open Access, the author has applied a CC BY public copyright licence to any Author Accepted Manuscript version arising from this submission.

\appendix

\section{Continuity equations in ADM form}\label{sec:appendix}

Here we write down the conservation law we use in terms of 3+1 decomposed ADM variables for a generic $\zeta^\mu$ vector.
We schematically write this down (see Eq.~(15) as \cite{Clough:2021qlv}) as
\begin{equation}\label{eqn:scheme}
    \frac{d}{dt} \int_{\Sigma} \dd^3 x \sqrt{\gamma} ~\mathcal{Q} = \\
\int_\Sigma \dd^3 x ~ \sqrt{\gamma} ~ \mathcal{S} + \\
\int_{\partial\Sigma} \sqrt{\sigma} \dd^2 x ~ N_\mu \mathcal{F}^\mu\,, 
\end{equation}
where we have made the following definitions
\begin{align}
    \mathcal{Q} = n_\mu \overline{\zeta}^\nu T^\mu{}_\nu\,\nonumber,\\
    \mathcal{S} = \alpha \nabla_\mu\overline{\zeta}^\nu T^\mu{}_\nu\,\nonumber,\\
    \mathcal{F^\mu} = \left(0, \alpha ~\overline{\zeta}^\nu T^i{}_\nu\right)\,.
    \label{eqn:QSF}
\end{align}
We then use the definition of the normal unit vector, $n_\mu=\left(-\alpha, 0,0,0\right)$, and the 3+1 decomposition of the stress-energy tensor,
\begin{equation}
    T^\mu{}_\nu = \rho n^\mu n_\nu + S^\mu n_\nu + n^\mu S_\nu + S^\mu{}_\nu\,,
\end{equation}
to write \eqref{eqn:QSF} as
\begin{align}
\mathcal{Q} &= -\alpha \rho \overline{\zeta}^0 + \left( \beta_j \overline{\zeta}^0 + \gamma_{ij}\overline{\zeta}^i \right) S^j\,,\nonumber\\
\mathcal{F}^i &= \overline{\zeta}^0\left( \beta^i \alpha \rho - \alpha^2S^i - \beta^i \beta^j S_j  + \alpha \beta^k S^i{}_k \right) \nonumber \\
        &\quad+ \left( \alpha S^i{}_j\overline{\zeta}^j-\beta^iS_k\overline{\zeta}^k \right)\,,\nonumber\\
\mathcal{S} &= \left( \alpha \rho \beta^i - \alpha^2 S^i - \beta^i \beta^k S_k + \alpha \beta^kS^i{}_k \right) \partial_i \overline{\zeta}^0 \nonumber \\
&\quad+ \left( \alpha S^i{}_j - S_j \beta^i \right)\partial_i \overline{\zeta}^j \nonumber \\
&\quad+ \overline{\zeta}^0\left( -\rho \partial_t \alpha + S_i\partial_t \beta^i + \tfrac{\alpha}2 S^{ij} \partial_t \gamma_{ij}\right) \nonumber \\
&\quad+\overline{\zeta}^i \left( -\rho\partial_i\alpha + S_j\partial_i\beta^j+\alpha S^k{}_j {}^{(3)}\Gamma^j_{ik} \right)\,.\label{eqn:QFS3D}
\end{align}
Note that these expressions reduce to Eq.~(10), (11), and (18) in \cite{Clough:2021qlv} with $\overline{\zeta}^\mu = (1,0,0,0)$, and Eq.~(12), (13), and (19) in \cite{Clough:2021qlv} with $\overline{\zeta}^\mu = \delta^\mu_i$.

These equations are generalizations of Eq.~(18) and (19) in \cite{Clough:2021qlv}. In the current version of the code, we implement \eqref{eqn:QFS3D} with the ADM quantities from the metric. It is equivalent to \eqref{eq:summary} in the main text, but in a form more suitable for use in general-relativistic simulations where the metric is evolved.

\section{Convergence test}\label{sec:appendix2}

\begin{figure}[h]
\centering
\includegraphics[width=0.48\textwidth]{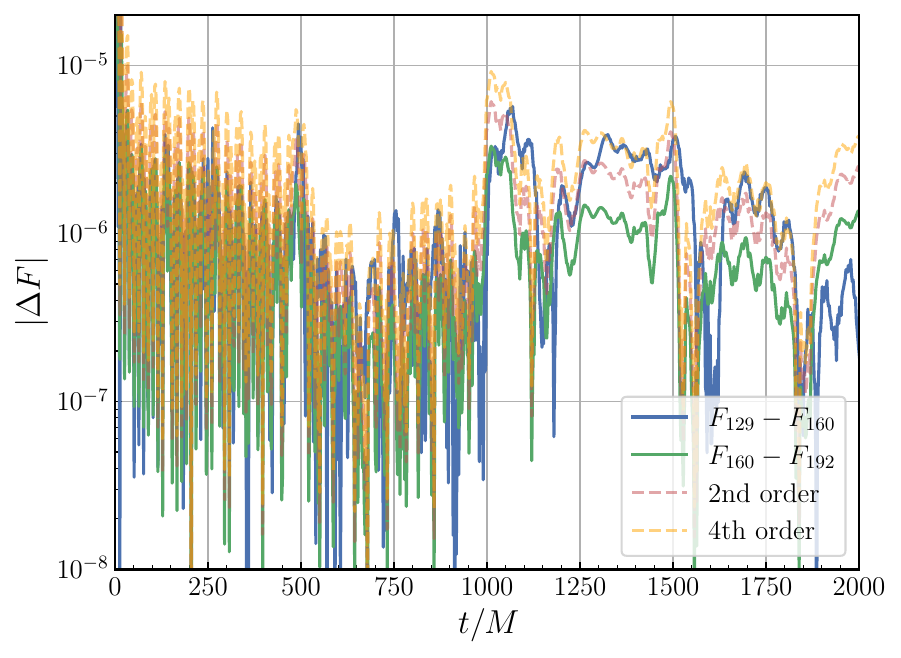}
\caption{Test of convergence for the extracted dynamical friction force. The blue solid line shows the absolute value of the difference of the dynamical friction force, $|\Delta F_1|$ between the base run with $N_1=128$ and a middle-resolution run with $N_2 = 160$. The green solid line shows that difference between the middle-resolution run and high-resolution run with $N_3=192$, $|\Delta F_2|$. The red and the orange dashed lines show the expected $|\Delta F_1|$ based on second order and fourth order convergence, respectively.}\label{fig:convergence}
\end{figure}

We perform a convergence test using the parameters $\alpha/M = 0.05$, $r/M = 900$, and $v=0.6$. All simulations have a box size of $L = 4096M$ and 9 refinement levels, as reported in the main text. We use three different resolutions, $N_1 = 128$ (the resolution used for results in the main text), $N_2 = 160$, and $N_3 = 192$. We show the absolute difference of the extracted force between runs with different resolutions in Fig.~\ref{fig:convergence}.

The difference of the extracted dynamical friction force between the mid- and the high-resolution simulations are approximately consistent with fourth-order convergence up to $t\sim 1800M$. Since $L=4096$, this is roughly consistent with the time where boundary effects propagate to the center of the computational grid, spoiling the convergence. 
Nonetheless, the magnitude of the difference is three orders of magnitude smaller than the dynamical friction force. At later times in the simulations, the value of the force oscillates around the constant value, and the amplitude of those oscillations decreases with time. Therefore the extracted dynamical friction force is robust to these late boundary effects.

%

\newpage

\bibliography{mybib}

@article{Hui:2016ltb,
    author = "Hui, Lam and Ostriker, Jeremiah P. and Tremaine, Scott and Witten, Edward",
    title = "{Ultralight scalars as cosmological dark matter}",
    eprint = "1610.08297",
    archivePrefix = "arXiv",
    primaryClass = "astro-ph.CO",
    doi = "10.1103/PhysRevD.95.043541",
    journal = "Phys. Rev. D",
    volume = "95",
    number = "4",
    pages = "043541",
    year = "2017"
}

@article{Barausse:2007ph,
      author         = "Barausse, Enrico",
      title          = "{Relativistic dynamical friction in a collisional fluid}",
      journal        = "Mon. Not. Roy. Astron. Soc.",
      volume         = "382",
      year           = "2007",
      pages          = "826-834",
      doi            = "10.1111/j.1365-2966.2007.12408.x",
      eprint         = "0709.0211",
      archivePrefix  = "arXiv",
      primaryClass   = "astro-ph",
      SLACcitation   = "%%CITATION = ARXIV:0709.0211;%%"
}

@article{Ostriker:1998fa,
      author         = "Ostriker, Eve C.",
      title          = "{Dynamical friction in a gaseous medium}",
      journal        = "Astrophys. J.",
      volume         = "513",
      year           = "1999",
      pages          = "252",
      doi            = "10.1086/306858",
      eprint         = "astro-ph/9810324",
      archivePrefix  = "arXiv",
      primaryClass   = "astro-ph",
      SLACcitation   = "%%CITATION = ASTRO-PH/9810324;%%"
}

@article{Syer:1994vr,
      author         = "Syer, D.",
      title          = "{Relativistic dynamical friction in the weak scattering
                        limit}",
      journal        = "Mon. Not. Roy. Astron. Soc.",
      volume         = "270",
      year           = "1994",
      pages          = "205",
      doi            = "10.1093/mnras/270.1.205",
      eprint         = "astro-ph/9404063",
      archivePrefix  = "arXiv",
      primaryClass   = "astro-ph",
      SLACcitation   = "%%CITATION = ASTRO-PH/9404063;%%"
}

@article{Hui:2021tkt,
    author = "Hui, Lam",
    title = "{Wave Dark Matter}",
    eprint = "2101.11735",
    archivePrefix = "arXiv",
    primaryClass = "astro-ph.CO",
    doi = "10.1146/annurev-astro-120920-010024",
    journal = "Ann. Rev. Astron. Astrophys.",
    volume = "59",
    pages = "247--289",
    year = "2021"
}

@ARTICLE{Tremaine:1984,
       author = {{Tremaine}, S. and {Weinberg}, M.~D.},
        title = "{Dynamical friction in spherical systems.}",
      journal = "Mon. Not. Roy. Astron. Soc.",
         year = 1984,
        month = aug,
       volume = {209},
        pages = {729-757},
          doi = {10.1093/mnras/209.4.729},
       adsurl = {https://ui.adsabs.harvard.edu/abs/1984MNRAS.209..729T}
}

@article{Muto:2011qv,
    author = "Muto, Takayuki and Takeuchi, Taku and Ida, Shigeru",
    title = "{On the Interaction between a Protoplanetary Disk and a Planet in an Eccentric Orbit: Application of Dynamical Friction}",
    eprint = "1106.0417",
    archivePrefix = "arXiv",
    primaryClass = "astro-ph.EP",
    doi = "10.1088/0004-637X/737/1/37",
    journal = "Astrophys. J.",
    volume = "737",
    pages = "37",
    year = "2011"
}

@article{Correia:2022gcs,
    author = "Correia, Miguel",
    title = "{Covariant formulation of relativistic mechanics}",
    eprint = "2202.04658",
    archivePrefix = "arXiv",
    primaryClass = "gr-qc",
    doi = "10.1103/PhysRevD.105.084041",
    journal = "Phys. Rev. D",
    volume = "105",
    number = "8",
    pages = "084041",
    year = "2022"
}

@article{Chandrasekhar:1943ys,
    author = "Chandrasekhar, Subrahmanyan",
    title = "{Dynamical Friction. I. General Considerations: the Coefficient of Dynamical Friction}",
    doi = "10.1086/144517",
    journal = "Astrophys. J.",
    volume = "97",
    pages = "255",
    year = "1943"
}

@article{Annulli:2020ilw,
    author = "Annulli, Lorenzo and Cardoso, Vitor and Vicente, Rodrigo",
    title = "{Stirred and shaken: Dynamical behavior of boson stars and dark matter cores}",
    eprint = "2007.03700",
    archivePrefix = "arXiv",
    primaryClass = "astro-ph.HE",
    doi = "10.1016/j.physletb.2020.135944",
    journal = "Phys. Lett. B",
    volume = "811",
    pages = "135944",
    year = "2020"
}

@article{Annulli:2020lyc,
    author = "Annulli, Lorenzo and Cardoso, Vitor and Vicente, Rodrigo",
    title = "{Response of ultralight dark matter to supermassive black holes and binaries}",
    eprint = "2009.00012",
    archivePrefix = "arXiv",
    primaryClass = "gr-qc",
    doi = "10.1103/PhysRevD.102.063022",
    journal = "Phys. Rev. D",
    volume = "102",
    number = "6",
    pages = "063022",
    year = "2020"
}

@article{Hartman:2020fbg,
    author = "Hartman, S. T. H. and Winther, H. A. and Mota, D. F.",
    title = "{Dynamical friction in Bose-Einstein condensed self-interacting dark matter at finite temperatures, and the Fornax dwarf spheroidal}",
    eprint = "2011.00116",
    archivePrefix = "arXiv",
    primaryClass = "astro-ph.CO",
    doi = "10.1051/0004-6361/202039865",
    journal = "Astron. Astrophys.",
    volume = "647",
    pages = "A70",
    year = "2021"
}

@ARTICLE{Ruderman:1971,
       author = {{Ruderman}, M.~A. and {Spiegel}, E.~A.},
        title = "{Galactic Wakes}",
      journal = {\apj},
         year = 1971,
        month = apr,
       volume = {165},
        pages = {1},
          doi = {10.1086/150870},
       adsurl = {https://ui.adsabs.harvard.edu/abs/1971ApJ...165....1R}
}

@ARTICLE{Petrich:1989,
       author = {{Petrich}, Loren I. and {Shapiro}, Stuart L. and {Stark}, Richard F. and {Teukolsky}, Saul A.},
        title = "{Accretion onto a Moving Black Hole: A Fully Relativistic Treatment}",
      journal = {\apj},
         year = 1989,
        month = jan,
       volume = {336},
        pages = {313},
          doi = {10.1086/167013},
       adsurl = {https://ui.adsabs.harvard.edu/abs/1989ApJ...336..313P}
}

@ARTICLE{Rephaeli:1980,
       author = {{Rephaeli}, Y. and {Salpeter}, E.~E.},
        title = "{Flow past a massive object and the gravitational drag}",
      journal = {\apj},
         year = 1980,
        month = aug,
       volume = {240},
        pages = {20-24},
          doi = {10.1086/158202},
       adsurl = {https://ui.adsabs.harvard.edu/abs/1980ApJ...240...20R}
}

@article{Lancaster:2019mde,
    author = "Lancaster, Lachlan and Giovanetti, Cara and Mocz, Philip and Kahn, Yonatan and Lisanti, Mariangela and Spergel, David N.",
    title = "{Dynamical Friction in a Fuzzy Dark Matter Universe}",
    eprint = "1909.06381",
    archivePrefix = "arXiv",
    primaryClass = "astro-ph.CO",
    doi = "10.1088/1475-7516/2020/01/001",
    journal = "JCAP",
    volume = {1},
    pages = {1},
    year = {2020}
}

@ARTICLE{BarOr:2019,
       author = {{Bar-Or}, Ben and {Fouvry}, Jean-Baptiste and {Tremaine}, Scott},
        title = "{Relaxation in a Fuzzy Dark Matter Halo}",
      journal = {\apj},
         year = 2019,
        month = jan,
       volume = {871},
       number = {1},
          eid = {28},
        pages = {28},
          doi = {10.3847/1538-4357/aaf28c},
archivePrefix = {arXiv},
       eprint = {1809.07673},
 primaryClass = {astro-ph.GA},
       adsurl = {https://ui.adsabs.harvard.edu/abs/2019ApJ...871...28B}
}

@article{Wang:2021udl,
    author = "Wang, Yourong and Easther, Richard",
    title = "{Dynamical friction from ultralight dark matter}",
    eprint = "2110.03428",
    archivePrefix = "arXiv",
    primaryClass = "gr-qc",
    doi = "10.1103/PhysRevD.105.063523",
    journal = "Phys. Rev. D",
    volume = "105",
    number = "6",
    pages = "063523",
    year = "2022"
}

@ARTICLE{Chowdhury:2021,
       author = {{Dutta Chowdhury}, Dhruba and {van den Bosch}, Frank C. and {Robles}, Victor H. and {van Dokkum}, Pieter and {Schive}, Hsi-Yu and {Chiueh}, Tzihong and {Broadhurst}, Tom},
        title = "{On the Random Motion of Nuclear Objects in a Fuzzy Dark Matter Halo}",
      journal = {\apj},
         year = 2021,
        month = jul,
       volume = {916},
       number = {1},
          eid = {27},
        pages = {27},
          doi = {10.3847/1538-4357/ac043f},
archivePrefix = {arXiv},
       eprint = {2105.05268},
 primaryClass = {astro-ph.GA},
       adsurl = {https://ui.adsabs.harvard.edu/abs/2021ApJ...916...27D}
}

@article{Boudon:2022dxi,
    author = "Boudon, Alexis and Brax, Philippe and Valageas, Patrick",
    title = "{Subsonic accretion and dynamical friction for a black hole moving through a self-interacting scalar dark matter cloud}",
    eprint = "2204.09401",
    archivePrefix = "arXiv",
    primaryClass = "astro-ph.CO",
    doi = "10.1103/PhysRevD.106.043507",
    journal = "Phys. Rev. D",
    volume = "106",
    number = "4",
    pages = "043507",
    year = "2022"
}

@article{Clough:2021qlv,
    author = "Clough, Katy",
    title = "{Continuity equations for general matter: applications in numerical relativity}",
    eprint = "2104.13420",
    archivePrefix = "arXiv",
    primaryClass = "gr-qc",
    doi = "10.1088/1361-6382/ac10ee",
    journal = "Class. Quant. Grav.",
    volume = "38",
    number = "16",
    pages = "167001",
    year = "2021"
}

@article{Vicente:2019ilr,
    author = "Vicente, Rodrigo and Cardoso, Vitor and Zilh\~ao, Miguel",
    title = "{Dynamical friction in slab geometries and accretion disks}",
    eprint = "1905.06353",
    archivePrefix = "arXiv",
    primaryClass = "astro-ph.GA",
    doi = "10.1093/mnras/stz2526",
    journal = "Mon. Not. Roy. Astron. Soc.",
    volume = "489",
    number = "4",
    pages = "5424--5435",
    year = "2019"
}

@article{Traykova:2021dua,
    author = "Traykova, Dina and Clough, Katy and Helfer, Thomas and Berti, Emanuele and Ferreira, Pedro G. and Hui, Lam",
    title = "{Dynamical friction from scalar dark matter in the relativistic regime}",
    eprint = "2106.08280",
    archivePrefix = "arXiv",
    primaryClass = "gr-qc",
    doi = "10.1103/PhysRevD.104.103014",
    journal = "Phys. Rev. D",
    volume = "104",
    number = "10",
    pages = "103014",
    year = "2021"
}

@article{Vicente:2022ivh,
    author = "Vicente, Rodrigo and Cardoso, Vitor",
    title = "{Dynamical friction of black holes in ultralight dark matter}",
    eprint = "2201.08854",
    archivePrefix = "arXiv",
    primaryClass = "gr-qc",
    doi = "10.1103/PhysRevD.105.083008",
    journal = "Phys. Rev. D",
    volume = "105",
    number = "8",
    pages = "083008",
    year = "2022"
}

@article{Buehler:2022tmr,
    author = "Buehler, Robin and Desjacques, Vincent",
    title = "{Dynamical friction in fuzzy dark matter: Circular orbits}",
    eprint = "2207.13740",
    archivePrefix = "arXiv",
    primaryClass = "astro-ph.CO",
    doi = "10.1103/PhysRevD.107.023516",
    journal = "Phys. Rev. D",
    volume = "107",
    number = "2",
    pages = "023516",
    year = "2023"
}

@article{Tomaselli:2023ysb,
    author = "Tomaselli, Giovanni Maria and Spieksma, Thomas F. M. and Bertone, Gianfranco",
    title = "{Dynamical friction in gravitational atoms}",
    eprint = "2305.15460",
    archivePrefix = "arXiv",
    primaryClass = "gr-qc",
    doi = "10.1088/1475-7516/2023/07/070",
    journal = "JCAP",
    volume = "07",
    pages = "070",
    year = "2023"
}

@article{Traykova:2023qyv,
    author = "Traykova, Dina and Vicente, Rodrigo and Clough, Katy and Helfer, Thomas and Berti, Emanuele and Ferreira, Pedro G. and Hui, Lam",
    title = "{Relativistic drag forces on black holes from scalar dark matter clouds of all sizes}",
    eprint = "2305.10492",
    archivePrefix = "arXiv",
    primaryClass = "gr-qc",
    doi = "10.1103/PhysRevD.108.L121502",
    journal = "Phys. Rev. D",
    volume = "108",
    number = "12",
    pages = "L121502",
    year = "2023"
}

@article{Wang:2024cej,
    author = "Wang, Zipeng and Helfer, Thomas and Traykova, Dina and Clough, Katy and Berti, Emanuele",
    title = "{Gravitational Magnus effect from scalar dark matter}",
    eprint = "2402.07977",
    archivePrefix = "arXiv",
    primaryClass = "gr-qc",
    doi = "10.1103/PhysRevD.110.024009",
    journal = "Phys. Rev. D",
    volume = "110",
    number = "2",
    pages = "024009",
    year = "2024"
}

@book{Landau:1975pou,
    author = "Landau, L. D. and Lifschits, E. M.",
    title = "{The Classical Theory of Fields}",
    doi = "10.1016/c2009-0-14608-1",
    isbn = "978-0-08-018176-9",
    publisher = "Pergamon Press",
    address = "Oxford",
    series = "Course of Theoretical Physics",
    volume = "Volume 2",
    year = "1975"
}

@article{Aurrekoetxea:2023jwk,
    author = "Aurrekoetxea, Josu C. and Clough, Katy and Bamber, Jamie and Ferreira, Pedro G.",
    title = "{Effect of Wave Dark Matter on Equal Mass Black Hole Mergers}",
    eprint = "2311.18156",
    archivePrefix = "arXiv",
    primaryClass = "gr-qc",
    doi = "10.1103/PhysRevLett.132.211401",
    journal = "Phys. Rev. Lett.",
    volume = "132",
    number = "21",
    pages = "211401",
    year = "2024"
}

@article{Roy:2025qaa,
    author = "Roy, Soumen and Vicente, Rodrigo and Aurrekoetxea, Josu C. and Clough, Katy and Ferreira, Pedro G.",
    title = "{Scalar Fields around Black Hole Binaries in LIGO-Virgo-KAGRA}",
    eprint = "2510.17967",
    archivePrefix = "arXiv",
    primaryClass = "gr-qc",
    reportNumber = "MIT-CTP/5937",
    doi = "10.1103/fv9z-zkxx",
    journal = "Phys. Rev. Lett.",
    volume = "136",
    number = "19",
    pages = "191402",
    year = "2026"
}

\end{document}